\pdfoutput=1
\documentclass[11pt,twoside,a4paper,cmspaper,final,collab]{cms-tdr}
\def\svnVersion{c3e0fc6}\def\svnDate{2022/06/30}\def\cmsCernNoTag{CERN-EP-2021-238}\def\cmsCernDate{\today}\def\cmsMessage{Published in Physics Letters B as \href{http://dx.doi.org/10.1016/j.physletb.2022.137263}{\doi{10.1016/j.physletb.2022.137263}.}}
\begin{document}\cmsNoteHeader{EXO-20-007}

\newcommand{\ROne}{\ensuremath{\mathrm{R}_1}\xspace}
\newcommand{\RTwo}{\ensuremath{\mathrm{R}_2}\xspace}
\newcommand{\POne}{\ensuremath{\mathrm{P}_1}\xspace}
\newcommand{\PTwo}{\ensuremath{\mathrm{P}_2}\xspace}
\newcommand{\PThree}{\ensuremath{\mathrm{P}_3}\xspace}
\newcommand{\MROne}{\ensuremath{m\!\left(\ROne\right)}\xspace}
\newcommand{\MRTwo}{\ensuremath{m\!\left(\RTwo\right)}\xspace}
\newcommand{\Ratio}{\ensuremath{\rho_{m}}\xspace}

\newcommand{\mjj}{\ensuremath{m_{\mathrm{jj}}}\xspace}
\newcommand{\mjjThr}{\ensuremath{\mjj^{\text{thr}}}\xspace}
\newcommand{\mjjMin}{\ensuremath{\mjj^{\text{min}}}\xspace}
\newcommand{\Nsub}{\textit{N}-subjettiness\xspace}
\newcommand{\tauTO}{\ensuremath{\tau_{21}}\xspace}
\newcommand{\MRTwojet}{\ensuremath{m_{\mathrm{Rjet}}}\xspace}
\newcommand{\MPThreejet}{\ensuremath{m_{\mathrm{Pjet}}}\xspace}

\newcommand{\gkk}{\ensuremath{\mathrm{G}_{\mathrm{KK}}}\xspace}
\newcommand{\Mgkk}{\ensuremath{m\!\left(\gkk\right)}\xspace}
\newcommand{\Mphi}{\ensuremath{m\!\left(\phi\right)}\xspace}

\newcommand{\gravCoupling}{\ensuremath{g_{\text{grav}}}\xspace}
\newcommand{\gkkCoupling}{\ensuremath{g_{\mathrm{GKK}}}\xspace}

\ifthenelse{\boolean{cms@external}}{\providecommand{\cmsLeft}{upper\xspace}}{\providecommand{\cmsLeft}{left\xspace}}
\ifthenelse{\boolean{cms@external}}{\providecommand{\cmsRight}{lower\xspace}}{\providecommand{\cmsRight}{right\xspace}}

\cmsNoteHeader{EXO-20-007}
\title{Search for high-mass resonances decaying to a jet and a Lorentz-boosted resonance in proton-proton collisions at  \texorpdfstring{$\sqrt{s}=13\TeV$}{sqrt(s) = 13 TeV}}

\date{\today}

\abstract{
A search is reported for high-mass hadronic resonances that decay to a parton and a Lorentz-boosted resonance, which in turn decays into a pair of partons. The search is based on data collected with the CMS detector at the LHC in proton-proton collisions at $\sqrt{s}=13\TeV$, corresponding to an integrated luminosity of 138\fbinv. The boosted resonance is reconstructed as a single wide jet with substructure consistent with a two-body decay. The high-mass resonance is thus considered as a dijet system. The jet substructure information and the kinematic properties of cascade resonance decays are exploited to disentangle the signal from the large quantum chromodynamics multijet background. The dijet mass spectrum is analyzed for the presence of new high-mass resonances, and is found to be consistent with the standard model background predictions. Results are interpreted in a warped extra dimension model where the high-mass resonance is a Kaluza--Klein gluon, the boosted resonance is a radion, and the final state partons are all gluons. Limits on the production cross section are set as a function of the Kaluza--Klein gluon and radion masses. These limits exclude at 95\% confidence level models with Kaluza--Klein gluon masses in the range 2.0 to 4.3\TeV and radion masses in the range 0.20 to 0.74\TeV. By exploring a novel experimental signature, the observed limits on the Kaluza--Klein gluon mass are extended by up to about 1\TeV compared to previous searches.
}

\hypersetup{
pdfauthor={CMS Collaboration},
pdftitle={Search for high-mass resonances decaying to a jet and a Lorentz-boosted resonance in proton-proton collisions at sqrt(s) = 13 TeV},
pdfsubject={CMS},
pdfkeywords={CMS, search, exotica, trijet, resonance}}

\maketitle

\section{Introduction}

The inability of the Standard Model (SM) to address problems such as the large gap between the gravitational and electroweak energy scales and to provide an explanation for astronomical observations indicating the existence of dark matter~\cite{ARBEY2021103865} provides strong motivation for experimental searches for new physics. Many theories beyond the SM predict the existence of new particles that can be produced at colliders at the \TeVns energy scale.

Searches for hadronic resonances are particularly important at the CERN LHC, as any hypothetical particle produced via the strong interaction in proton-proton ($\Pp\Pp$) collisions can decay to quarks and gluons, which hadronize to form jets. Direct searches at the LHC have so far not found compelling evidence of new physics beyond the SM. The main background consists of SM quantum chromodynamics (QCD) processes that produce multiple jets in the final state (referred to as QCD multijet background in the following) and it is typically very large compared to the potential signals of new physics. If these new particles exist and are within the energy range of the LHC, they still could have been missed by experimental searches because they decay mainly into final-state configurations for which the current strategies have not been optimized.

\begin{figure}[ht]
	\centering
	\includegraphics[width=0.4\textwidth]{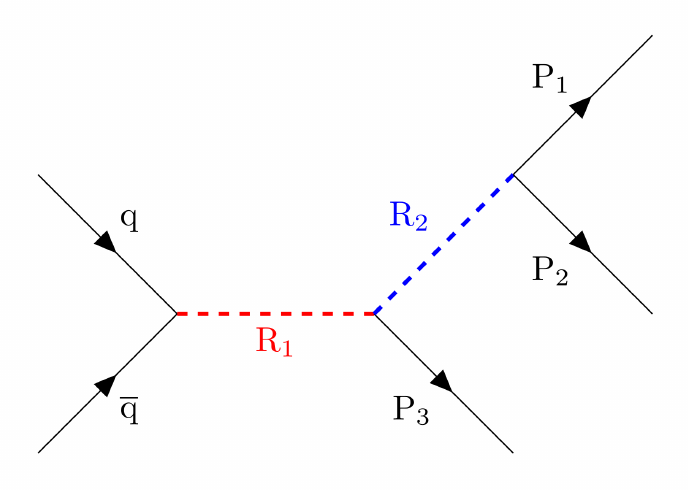}
	\caption{Feynman diagram of leading order production of the process $\ROne \to \RTwo + \PThree \to (\POne + \PTwo) + \PThree$ involving cascade decays of two new massive resonances \ROne and \RTwo to partons \POne, \PTwo, and \PThree in the final state.}
	\label{fig:diagram}
\end{figure}

The existing searches assume production of single resonances decaying to a pair of jets (dijet)~\cite{Sirunyan:2019vgj, Aad:2019hjw}, production of dijet resonances in association with an initial state radiation jet~\cite{CMS:2019mcu,CMS:2019emo,ATLAS:2018hbc}, photon~\cite{Sirunyan:2676241,ATLAS:2019itm} or lepton~\cite{Aad:2711475}, and pair production of resonances resulting in final states with four~\cite{ATLAS:2017jnp,Sirunyan:2018rlj} or more~\cite{CMS:2018pdq,ATLAS:2018umm,Sirunyan:2018duw} jets. This analysis extends those searches by considering a new process, where a resonance (\ROne) decays into a lighter resonance (\RTwo) and an SM particle (\PThree), $\PQq\PAQq \to \ROne \to \RTwo + \PThree \to (\POne + \PTwo) + \PThree$, as shown in Fig.~\ref{fig:diagram}. Such cascade resonance decays are foreseen by theoretical models beyond the SM that predict the existence of extra spatial dimensions~\cite{Agashe:2016kfr,Agashe:2017wss,Agashe:2020wph} or the existence of heavy partners of SM quarks~\cite{Redi:2013eaa}. We consider the case where $\mathrm{P}_x$ (with $x=1,2,3$) are all partons (quarks, antiquarks, or gluons, depending on the theoretical model considered).

{\tolerance=950 The experimental signature is characterized in the final state by the mass ratio $\Ratio=\MRTwo/\MROne$, where \MROne and \MRTwo are the masses of the two resonances. If \MRTwo is significantly smaller than \MROne, \RTwo is produced with large momentum, and its decay products are collimated and can be reconstructed as a single jet in the detector. The analysis presented here targets final states with two high-momentum reconstructed wide jets: the first jet (\PThree jet) comes from the hadronization of parton \PThree, and the second jet (\RTwo jet) contains the hadronization products of both \POne and \PTwo. This analysis considers scenarios where $\Ratio < 0.2$, to allow a sufficiently large boost of the \RTwo resonance, and is sensitive to \ROne resonance masses $\MROne > 2\TeV$. Scenarios with larger \Ratio values, where three well-separated jets from the \ROne decay are reconstructed, require a different analysis strategy, and are not discussed here.\par}

The analysis uses distributions of the dijet mass (\mjj), the invariant mass of the two reconstructed jets, and searches for a peak from the resonance \ROne. The latest searches for high-mass dijet resonances~\cite{Sirunyan:2019vgj, Aad:2019hjw} are sensitive to this final state, but they are not optimized for this particular cascade resonance decay. To increase the analysis sensitivity, we exploit the pattern of the particles inside these jets to distinguish between the signal, containing a massive \RTwo jet, and the main background from QCD multijet events that originate from hadronization of single partons. 

We consider a signal benchmark model with a warped extra dimension~\cite{Agashe:2016kfr} where \ROne is a spin 1 Kaluza--Klein gluon (\gkk) produced through the \textit{s} channel via quark-antiquark annihilation, \RTwo is a spin 0 radion ($\phi$), and \POne, \PTwo, and \PThree are all gluons. We assume that only the gluon field among the SM gauge fields is allowed to propagate in the entire bulk of the extra dimension~\cite{Agashe:2020wph}. Under this hypothesis, the \gkk can decay into a radion and a gluon, or into a quark-antiquark pair, and the radion only decays to a pair of gluons. The \ROne resonance is assumed to be narrow with a total decay width of about~1\% of its mass. The partial decay width of \gkk to a quark-antiquark pair scales as $(1/\gkkCoupling)^{2}$, while the partial decay width to a radion and a gluon is proportional to $(\gravCoupling/\gkkCoupling)^{2}$, as described in Ref.~\cite{Agashe:2016kfr}. Here \gravCoupling and \gkkCoupling are the gravitational and gauge couplings for \gkk, respectively, and are both free parameters of the theory. An increase of \gravCoupling enhances the branching fraction  of \ROne into the radion+gluon channel, while an increase of \gkkCoupling has the main effect of reducing the \gkk production cross section.  The two coupling values are estimated to be in the ranges $1 \lesssim \gravCoupling \lesssim 6$ and $3 \lesssim \gkkCoupling \lesssim 6$ using the theoretical assumptions discussed in Ref.~\cite{Agashe:2016kfr}.

As discussed above, in this model, the \gkk can decay into a radion and a gluon, or into a quark-antiquark pair. Existing bounds on the \gkk mass from $\gkk \to \PQq\PAQq$ decays are described in Section~\ref{sec:results} and compared with the results of this analysis. In the case where the \gkk decays to a radion and a gluon, in the model considered the radion can only decay into a pair of gluons, thus constraints on the radion mass from existing searches in all other final states do not apply. In addition, the CMS search for low-mass dijet resonances~\cite{CMS:2018mgb}, which studies the process $\Pg\Pg \to \phi \to \Pg\Pg$, is not sensitive in the range of radion masses for the particular choice of model couplings considered in this paper.

The analysis uses $\Pp\Pp$ collision data at a center-of-mass energy of 13\TeV collected with the CMS detector at the LHC in 2016, 2017, and 2018, corresponding to an integrated luminosity of 138\fbinv. Tabulated results are provided in the HEPData record for this analysis~\cite{hepdata}.

\section{The CMS detector}\label{sec:the_cms_detector}
The central feature of the CMS apparatus is a superconducting solenoid of 6\unit{m} internal diameter, providing a magnetic field of 3.8\unit{T}. Within the solenoid volume are a silicon pixel and strip tracker, a lead tungstate crystal electromagnetic calorimeter (ECAL), and a brass and scintillator hadron calorimeter (HCAL), each composed of a barrel and two endcap sections. Forward calorimeters extend the pseudorapidity coverage provided by the barrel and endcap detectors. Muons are measured in gas-ionization detectors embedded in the steel flux-return yoke outside the solenoid. A more detailed description of the CMS detector, together with a definition of the coordinate system used and the relevant kinematic variables, can be found in Ref.~\cite{Chatrchyan:2008zzk}.

Events of interest are selected using a two-tiered trigger system. The first level (L1), composed of custom hardware processors, uses information from the calorimeters and muon detectors to select events at a rate of around 100\unit{kHz} with a fixed latency of about 4\mus~\cite{CMS:2020cmk}. The second level, known as the high-level trigger (HLT), consists of a farm of processors running a version of the full event reconstruction software optimized for fast processing, and reduces the event rate to around 1\unit{kHz} before data storage~\cite{Khachatryan_2017}.

\section{Data sets and event selection}\label{sec:data_sets_and_event_selection}

Simulated signal samples are generated at leading order with \MGvATNLO v. 2.4.3~\cite{Alwall:2014mdg} for $\Ratio=0.1$ and 0.2, and with \gkk masses, \Mgkk, between 2 and 9\TeV in 1\TeV steps. The specific choice of coupling parameters \gravCoupling and \gkkCoupling used in the generation does not affect the decay kinematic distributions but only modifies the signal cross section. For this reason, the signal selection efficiencies and distributions of kinematic observables estimated using the simulated samples are valid also for models with different coupling parameters.

Simulations of the QCD multijet background are produced with the \PYTHIA 8.205~\cite{Sj_strand_2015} program. We use the QCD background simulated samples for the optimization of the analysis strategy, while the final background estimation is obtained through a fit to the dijet mass distributions in the data. 

Both the signal and background samples are generated using the next-to-next-to-leading (NNLO) order parton distribution function (PDF) set NNPDF3.1~\cite{NNPDF:2017mvq}. Fragmentation and hadronization are simulated with \PYTHIA 8.205~\cite{Sj_strand_2015} with the CP5~\cite{CMS:2019csb} underlying event tune. All simulated samples are processed with the full \GEANTfour-based~\cite{Agostinelli:602040} simulation of the CMS detector and they are reconstructed with the same suite of programs used for collision data.

Events are reconstructed using the CMS particle-flow (PF)~\cite{Sirunyan:2017ulk} algorithm, which combines information from every subsystem of the CMS detector to reconstruct and identify individual particles (called PF candidates). Particles produced in additional collisions within the same bunch crossing (pileup) are suppressed by applying a weight to each PF candidate, calculated by the pileup-per-particle identification (PUPPI) algorithm~\cite{Bertolini:2014bba}. It has been shown that the PUPPI algorithm mitigates the effects of pileup in the measurement of jet observables~\cite{puppi2020}.

At the HLT stage of the trigger system described in Section~\ref{sec:the_cms_detector}, the PF candidates are clustered into jets using the \FASTJET package~\cite{Cacciari:2011ma} with the anti-\kt algorithm~\cite{Cacciari:2008gp} and a distance parameter $R=0.4$ (AK4 jets).
Single-jet triggers, selecting events with a jet that exceeds a predefined \pt threshold, are used. Triggers that require \HT to exceed a threshold are also used, where \HT is the scalar sum of \pt for all AK4 jets in the event with $\pt> 30\GeV$ and $\abs{\eta} <3.0$. The HLT requires $\HT > 900$ or 1050\GeV, depending on the data-taking period, or at least one jet reconstructed with an increased distance parameter of $R=0.8$ and $\pt>550\GeV$.

In the offline selection, in order to collect the decay products of \RTwo for \Ratio values up to about 0.2, reconstructed wide jets are formed using the anti-\kt algorithm with $R = 1.5$. Jets with large size collect more effectively hard-gluon radiation that may occur from the parton \PThree, improving the dijet mass resolution. Hence, we use wide jets with $R = 1.5$ for the reconstruction of both the \RTwo jet and the \PThree jet in our analysis. In the following, ``jets" refers to these wide jets. Jets are corrected as a function of their \pt and $\eta$ to match the observed detector response~\cite{CMS-PAS-JME-16-003}. Jet masses are reconstructed with the soft drop algorithm~\cite{Larkoski_2014} with parameters $\beta=0$, $z_{\text{cut}}=0.1$, and $R_{0}=1.5$.

For each event we select jets with $\pt> 100\GeV$ and $\abs{\eta} < 2.5$. The two jets with largest \pt are defined as the leading jets. The $\eta$ separation between the two leading jets is required to be $\abs{\Delta\eta_{\mathrm{jj}}} < 1.3$, as in previous searches for dijet resonances~\cite{Sirunyan:2232269}. This requirement maximizes the analysis sensitivity by suppressing the background of dijet events from QCD \textit{t}-channel production processes, while keeping good acceptance for signal events produced via the $s$ channel. The invariant mass of the two jets is required to be $\mjj> 1.6\TeV$ to ensure that the trigger is fully efficient for events passing the offline selection. These selections restrict the region of the measurement predominantly to the central region, which corresponds to $\abs{\eta} \lesssim 1.5$. The analysis of simulated samples shows that the signal efficiency for these kinematic requirements is between 40 and 50\% for all \MROne and \Ratio values considered.

Using jet substructure information, we identify the \RTwo jet as the leading jet with the lowest associated \Nsub ratio (\tauTO) value~\cite{Thaler:2010tr}, while the other leading jet is identified as the \PThree jet. Signal events show a resonance peak in the distributions of both the soft drop mass of the \RTwo jet (\MRTwojet) and the reconstructed mass of the \ROne resonance (\mjj). For all signal hypotheses investigated in this search, in about 30--35\% of the events the jet substructure algorithm incorrectly tags the single jet coming from \PThree hadronization as the \RTwo jet candidate. Therefore, for these events, there is a resonance peak in the distribution of the reconstructed \PThree jet mass (\MPThreejet) instead of \MRTwojet.

\section{Analysis strategy and optimization}\label{sec:analysis_strategy_and_optimization}

\begin{figure}[thb!]
	\centering
    \includegraphics[width=0.49\textwidth]{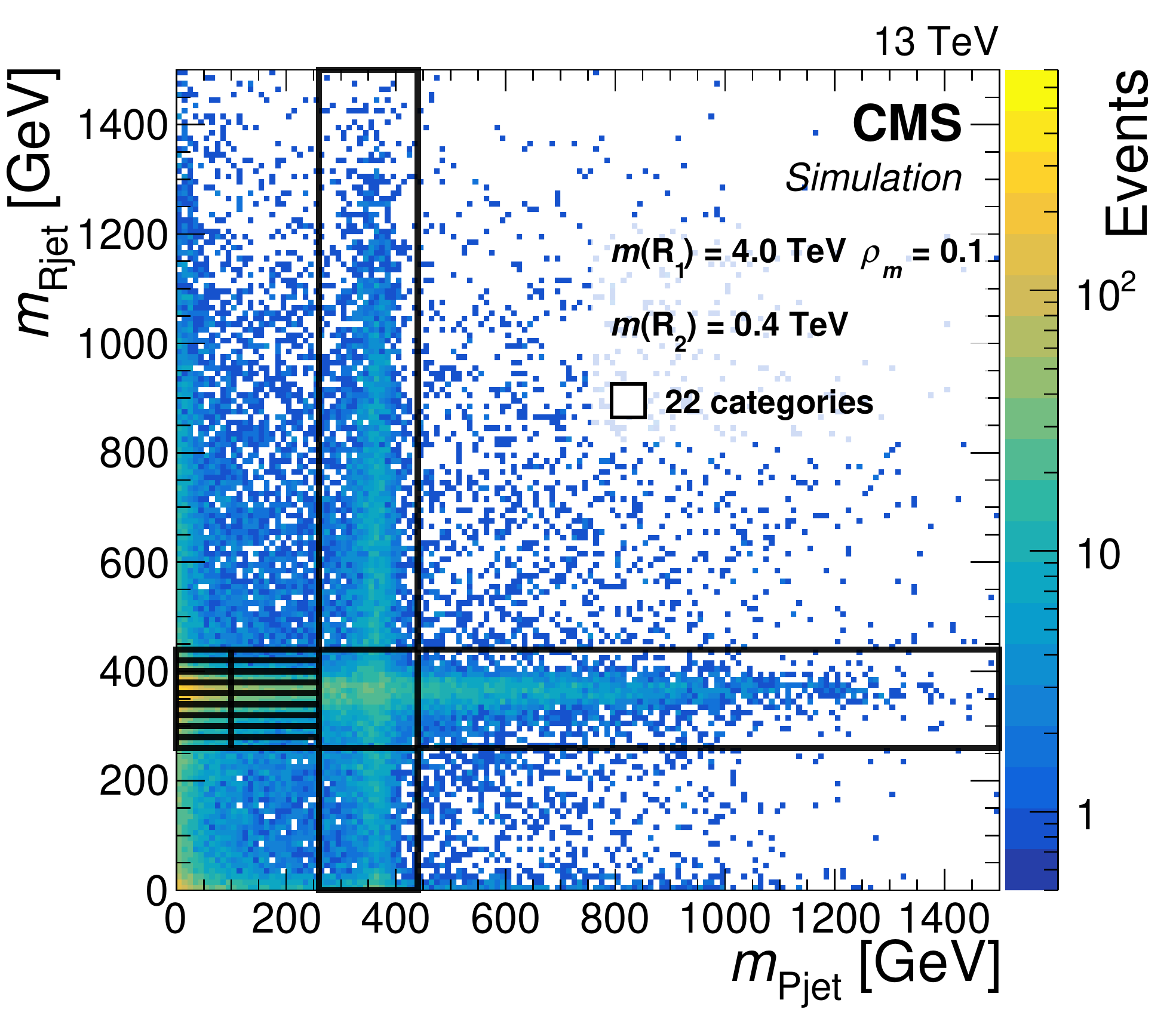}\hfill
    \includegraphics[width=0.49\textwidth]{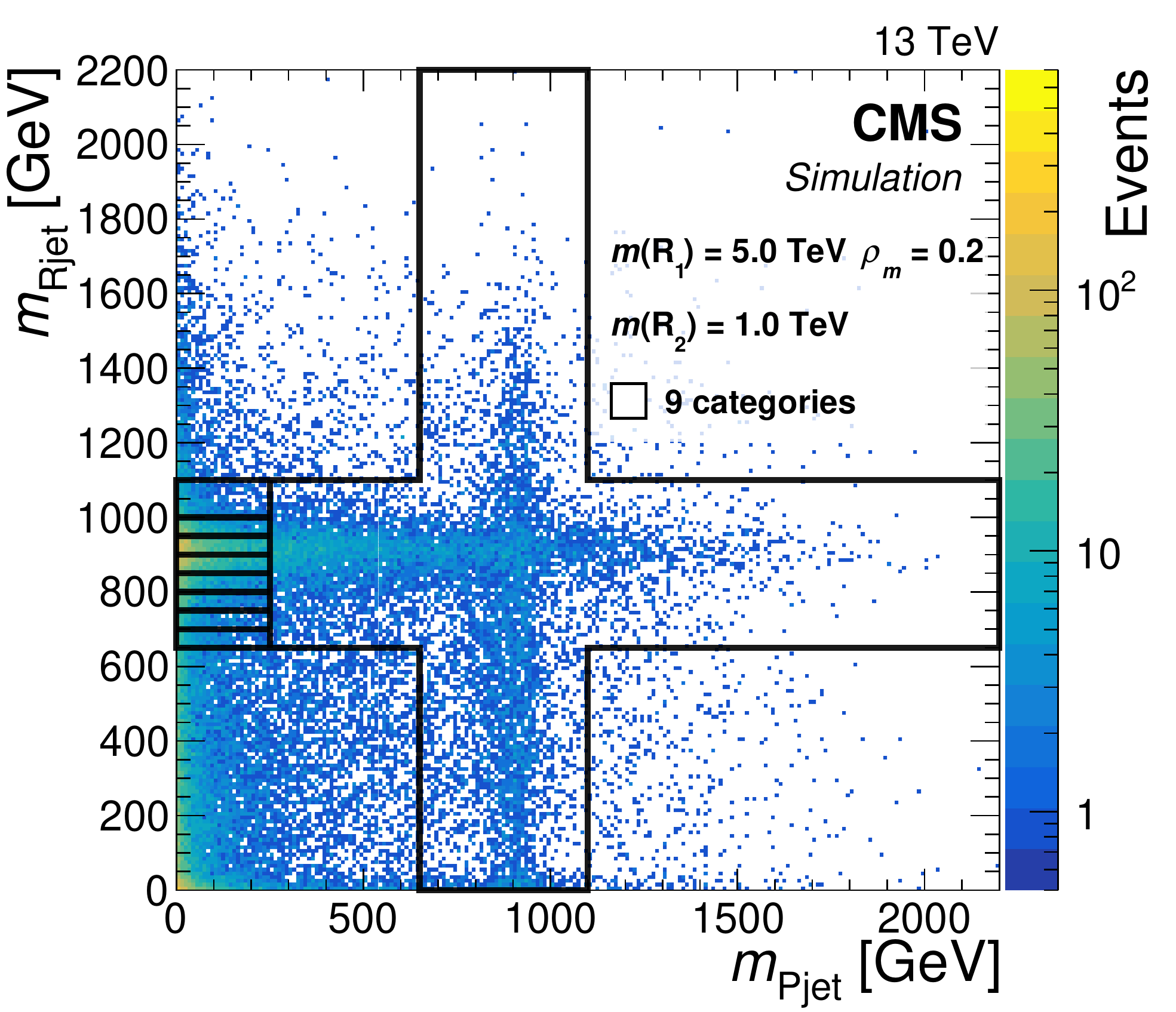}
    \caption{In the simulation, the reconstructed mass of the \RTwo jet candidate (\MRTwojet) vs. the reconstructed mass of the \PThree jet candidate (\MPThreejet) for \ROne resonance events originating from two different mass hypotheses. The \cmsLeft plot is for a \gkk with a mass $\Mgkk=\MROne=4\TeV$, decaying to a radion with a mass $\Mphi=\MRTwo=0.4\TeV$ and a gluon. The 22 event categories in this plane, within which the search in the dijet mass distribution is conducted, are shown with black boxes. The \cmsRight plot is for the same decay sequence, with masses $\Mgkk=5\TeV$ and $\Mphi=1\TeV$, for which the number of event categories is 9. For both plots, the cross-like shape is approximately centered on the second resonance pole mass \MRTwo for both the horizontal and the vertical axes.}
	\label{fig:categories}
\end{figure}

\begin{figure}[thb!]
	\centering
    \includegraphics[width=0.49\textwidth]{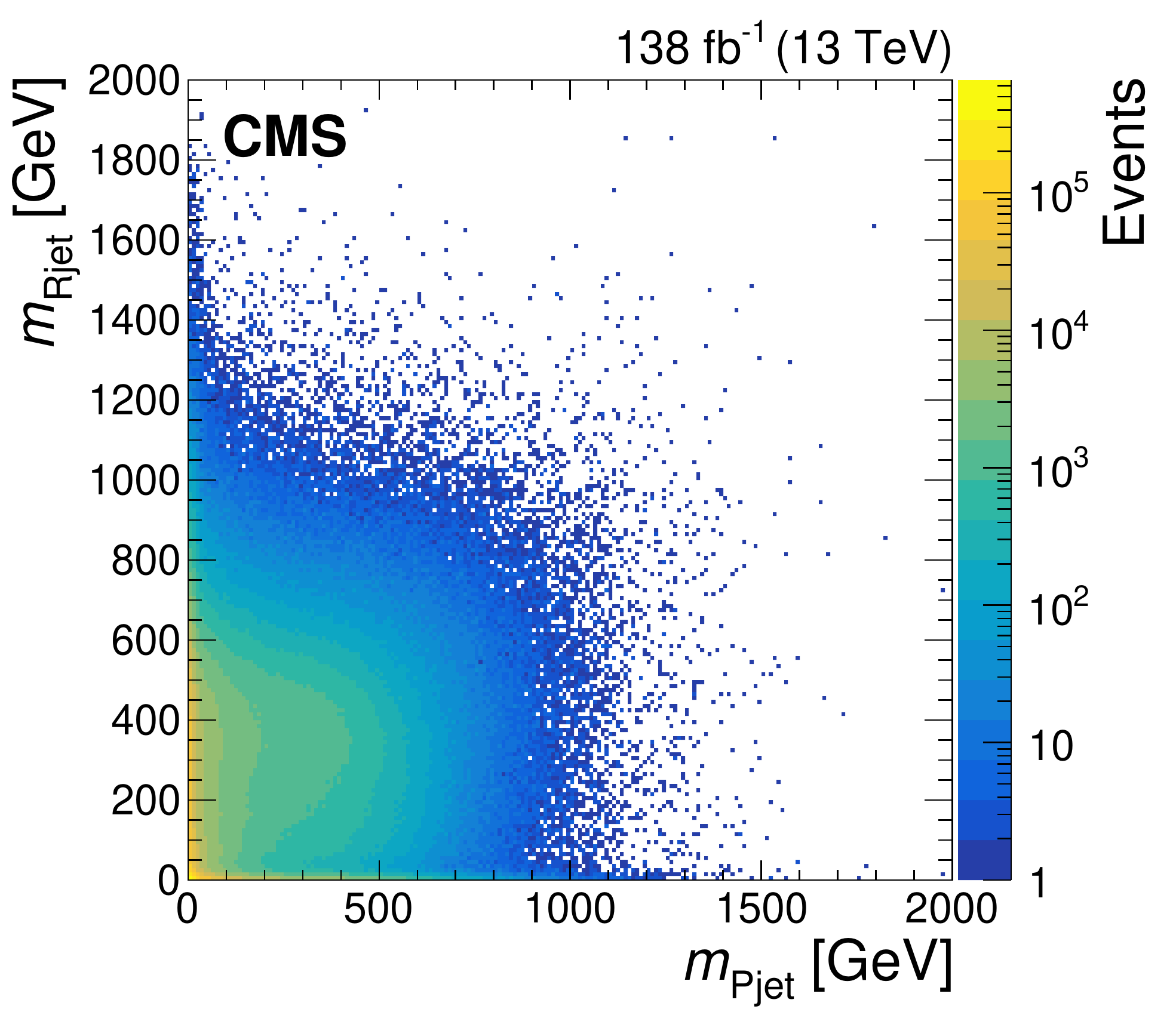}
    \caption{Distribution of the reconstructed mass of the \RTwo jet candidate (\MRTwojet) vs. the reconstructed mass of the \PThree jet candidate (\MPThreejet) for events in data, which are expected to arise primarily from QCD multijet events.}
    \label{fig:data_distribution}
\end{figure}

For signal events, we observe a characteristic cross-like shape in the \MRTwojet vs. \MPThreejet plane centered around the value of the \RTwo mass, where the horizontal (vertical) axis of the cross represents the events with correct (incorrect) \RTwo jet matching. Examples for two different signal hypotheses are illustrated in Fig.~\ref{fig:categories}. The data distribution in Fig.~\ref{fig:data_distribution}, dominated by QCD multijet background events, shows instead a smooth pattern in this two-dimensional jet mass plane, with a weak correlation between the two observables. 

The analysis strategy is to divide events into categories following the signal cross-like pattern, and to search for localized enhancements from the \ROne resonance in the \mjj distributions of each of these categories. The background decreases smoothly and rapidly with increasing dijet mass. The possible presence of a signal is investigated by fitting the observed dijet mass distribution with a function comprising both signal and background components. The background is modeled by a smooth monotonically decreasing function, and the signal is modeled by a function that describes the narrow resonance peak

To enhance the fit sensitivity and exploit all information from jet mass distributions, events are divided into the categories defined in the \MRTwojet vs. \MPThreejet plane as depicted in Fig.~\ref{fig:categories}. The boundaries of these categories are chosen to contain events from the \RTwo resonance, making a cross-like pattern centered on the value of \MRTwo considered, as shown in Fig.~\ref{fig:categories}. Only events with \MRTwojet and \MPThreejet values inside the cross are used in the analysis. The horizontal and vertical arms of the cross contain \MRTwojet and \MPThreejet values ranging from 65 to 110\% of \MRTwo for all \MPThreejet and \MRTwojet values, respectively. The window is asymmetric with respect to \MRTwo because the soft drop jet mass algorithm reconstructs a peak mass that is about 10\% lower than the nominal \RTwo mass. The window chosen optimizes the search sensitivity to a narrow resonance. Events in the cross are then more finely divided into multiple categories, with the number of categories decreasing as the mass of the \RTwo resonance increases. There are 22 categories when \MRTwo is less than 0.6\TeV, 9 categories when \MRTwo ranges from 0.6 to 1.2\TeV, and 1 category for \MRTwo greater than 1.2\TeV. The low \MPThreejet region of the horizontal arm of the cross is the region with the highest fraction of signal events and the largest background. We exploit the differences in \MRTwojet vs. \MPThreejet correlations between signal and background events to improve the analysis sensitivity. In the case of 22 categories, we divide the low \MPThreejet region of the horizontal arm of the cross into nine horizontal slices based on \MRTwojet, with a width approximately equal to the jet mass resolution (about 5\% of \MRTwo). As a result of the analysis optimization, each of these slices is further divided into two sub-categories, separating events with values of \MPThreejet below or above $0.25\MRTwo$. This approach allows us to exploit the line shape of the signal jet-mass distribution and to separate categories with a high signal-over-background ratio (near the \RTwo jet mass peak) from the other categories with lower sensitivity. The remaining region, corresponding to the vertical arm and the high \MPThreejet region of the horizontal arm of the cross, is divided into four categories as shown in the \cmsLeft plot of Fig.~\ref{fig:categories}. The jet mass range of these latter categories is wider in order to retain a sufficient number of events in data to perform the fit. At larger values of \MRTwo there are smaller numbers of events within the cross. To have event samples that are sufficiently large to ensure stable fits, the number of categories is first reduced to 9, as shown in the \cmsRight plot of Fig.~\ref{fig:categories}, and then to a single category corresponding to the entire cross. The improvement in sensitivity to new physics, evaluated as the relative reduction of the expected upper limits on signal cross section, is a factor between 1.5 and 2.5 (for \Ratio values between 0.175 and 0.1) compared to an inclusive analysis that has no event classification based on jet mass and jet substructure information. 

When testing signal hypotheses with higher \RTwo masses, the center of the cross formed by the categories in the  \MRTwojet vs. \MPThreejet plane shifts to higher values of the \RTwo and \PThree jet masses. Therefore, the jet masses are required to be higher and, consequently, the shape of the corresponding \mjj spectrum is modified by a turn-on effect that produces a broad peak. The resulting distribution cannot be modeled by the smoothly decreasing function describing the QCD background. To exclude this low dijet mass region from the analysis, a jet-mass-dependent minimum \mjj threshold (\mjjThr) is applied, which is specific to each category. The threshold is evaluated from simulated background samples. For each category, we compute the ratio between the \mjj spectra with and without the application of the selection on the \RTwo and \PThree jet masses. This ratio, as a function of \mjj, shows an increasing trend and a maximum before decreasing, and the same behavior is observed in both data and simulation. The \mjjThr is chosen to be 15\% higher than the position of the maximum. The chosen \mjjThr is the minimum value such that no significant bias is introduced in the signal extraction procedure described later.

\section{Background and signal model}\label{sec:background_and_signal_model}

A simultaneous binned maximum likelihood fit to the \mjj spectra of all categories is performed.
The bin size is a function of the dijet mass and approximately equal to the dijet mass resolution. The fit includes a signal and a background function. The SM background in a given category is modeled with an empirical three-parameter function $f(x) = p_0\left(  1-x \right)^{p_1} / x^{p_2}$, where $x=\mjj/\sqrt{s}$, which is a reparameterization of the function previously used in dijet resonance searches~\cite{Sirunyan:2019vgj}. For a given signal hypothesis, the total number of background parameters is therefore the number of categories multiplied by three. The signal shape of the \ROne resonance in each category is modeled with a double-sided Crystal Ball function~\cite{CrystalBallRef1,CrystalBallRef2}. All the parameters of the function describing the background are allowed to vary in the fit. With this approach the background estimation is obtained from data alone and does not depend on the simulation of QCD multijet events. The parameters of the signal function are determined from the simulated signal samples at different \MROne and \Ratio values. The resulting parameters are then linearly interpolated between \MROne and \Ratio points to obtain the intermediate signal shapes. The granularity of the interpolation is 100\GeV in \MROne and 0.0125 in \Ratio. The interpolation procedure has been tested and shown to provide realistic signal shapes. The resolution of the reconstructed signal peak is about 5\% of \MROne, for all the mass hypotheses considered. The parameter of interest is the modifier of the signal strength, which is a multiplicative factor of the signal normalization in each category, and is the same for all the categories.

We fit all \mjj spectra in the range $\mjjMin <\mjj< 1.25\MROne$, where \mjjMin is the greater of \mjjThr and $0.65\MROne$. If $\mjjThr>0.9\, \MROne$, the signal peak is truncated and the corresponding category is removed from the analysis to avoid signal biases in the fit. The total signal efficiency for events to pass the kinematic selection and be included in the fit range is usually between 20 and 30\%. In the region with $\Ratio \approx 0.2$ and $\MROne\lesssim 4\TeV$, the signal efficiency is approximately 10\%, because the signal peak is truncated as described above. Detailed signal injection tests show that the potential bias in the background prediction method is negligible for the entire range of signal hypotheses considered. The signal injection tests are performed as follows: pseudodata distributions are generated for a hypothesis including background but no signal, using the $f(x)$ background function, with parameters fixed to the values from the best fit to collision data. Pseudodata distributions are also produced including both the background and a signal, injected with a cross section equal to the 95\% confidence level (\CL) expected limit. These distributions are created for all signal hypotheses considered. Then, the fitting procedure is repeated for each pseudodata distribution, and the fitted signal cross section, along with its standard deviation, is obtained. We examine the distribution of the bias in units of standard deviations; that is, the difference between the injected signal cross section and the fitted signal cross section divided by the standard deviation of the fit. For all resonance masses, widths, and signal strengths considered, the
mean bias is less than one half a standard deviation, and in the vast majority of the cases it is
well below this criterion. In addition, these studies are performed with the pseudodata distributions generated from an alternative empirical function $f'(x) = p_0 \left(\exp\left(-p_1 x \right) \right) / x^{p_2}$ that also describes the data. This tests the flexibility of the $f(x)$ function to fit to a spectrum with a different shape.
The entire procedure described above is repeated for the alternative function, again yielding negligible biases. 

\section{Systematic uncertainties}\label{sec:systematic_uncertainties}

The dominant sources of systematic uncertainty are those related to the scale and resolution of the jet energy and jet mass, and to the \Nsub ratio \tauTO. The uncertainties in the jet energy scale and resolution translate, respectively, into uncertainties in the position and the width of the dijet mass shape for the signal. The effect of these uncertainties is propagated to the limits by shifting the dijet mass shape by $\pm$2\% and varying its reconstructed width by $\pm$20\%.
The uncertainties in the jet mass scale and resolution for jets with $R = 1.5$ have previously been evaluated in a CMS analysis~\cite{CMS:2019emo} that searched for Lorentz-boosted $\Pq\overline{\Pq}$ resonances in the mass range 40--450\GeV. The uncertainty in the \tauTO observable is obtained from a comparison between the \tauTO distributions in data and in simulated samples of QCD multijet processes, after applying the event selection described in Section~\ref{sec:data_sets_and_event_selection}. 
The uncertainties in the jet mass scale and resolution and in the \tauTO observable cause event migrations between categories, which translate into uncertainties in the signal normalization. These uncertainties are propagated to the limits by varying the signal normalization by a value that ranges between $\pm$1\% and $\pm$50\% of the central values obtained from the simulation, depending on the source of the uncertainty and on the category considered. The uncertainty on the integrated luminosity is 1.6\%~\cite{CMS-LUM-17-003,CMS-PAS-LUM-17-004,CMS-PAS-LUM-18-002} and it is propagated to the normalization of the signal.

A single dijet mass shape, the average over all categories for each value of \MROne and \Ratio, is used for the signal. This choice simplifies the fit procedure by avoiding large statistical uncertainties in the signal shape, and mildly affects the analysis sensitivity, resulting in a 10\% increase of the expected limit. The systematic uncertainty in the signal shape, from observed differences between the average signal shape and the shapes of each category, is estimated from simulated signal samples, and is 2\% in the peak position and 30\% in the width of the signal peak

The mean and the width of the Gaussian core of the signal Crystal Ball function, together with the signal efficiencies in each category, are treated as nuisance parameters in the fit, and are allowed to float within systematic uncertainties. The impact of all the sources of systematic uncertainties in the other parameters of the Crystal Ball function is negligible. Therefore, these other parameters are fixed to the values obtained from the simulated signal samples. 

The total effect of all the systematic uncertainties is to increase the upper limits on the signal cross section by up to 20\%.

\section{Results}\label{sec:results}

We test for the presence of \gkk signals for masses between 2 and 9\TeV and \Ratio values between 0.1 and 0.2. We are not sensitive to lower signal masses, because of the trigger and event selection criteria discussed above, and larger values of \Ratio are not considered because of the small signal efficiency. We find results compatible with background predictions. We compute the signal significance using the logarithm of the ratio of profile likelihoods as the test statistic. The distribution of the test statistic is obtained with a frequentist approach, using pseudodata samples with a large number of events. The most significant excess in the data, when interpreted as a signal with $\Mgkk = 2.9\TeV$ and $\Mphi = 0.4\TeV$, corresponds to a local significance of 3.2 standard deviations.

We evaluate the global significance of this excess by taking into account the look-elsewhere effect~\cite{LEE}. Since the categories defined in the \MRTwojet vs. \MPThreejet plane for the signal hypotheses, with similar \MRTwo, have a large overlap, we evaluate the look-elsewhere effect for a subset of signal hypotheses with $\Mphi= 400$, 840 and 1440\GeV, where \Mgkk ranges from 2 to 9\TeV.
These signal hypotheses correspond to three sets of categories and have minimal overlap in the \RTwo vs. \PThree jet mass plane. Therefore, they are considered to be independent samples of events. 
Considering only this subset of signal hypotheses, the global significance is found to be 1.8 standard deviations. For the full range of tested signal hypotheses, the global significance of the excess would be lower than this value.

\begin{figure}[ht]
    \centering
	\includegraphics[width=0.49\textwidth]{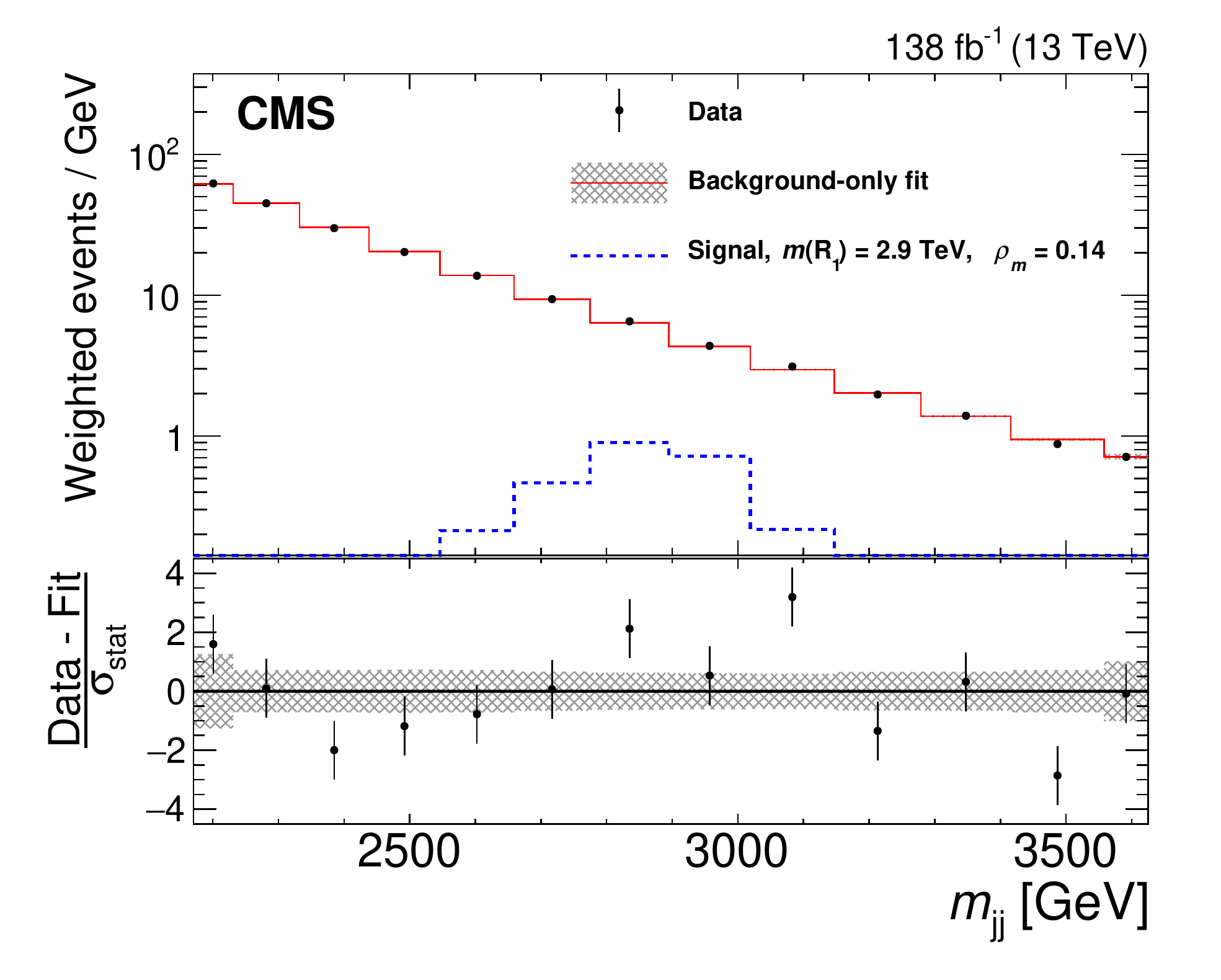} 
    \caption{Dijet mass spectrum, from the combination of the spectra within 22 categories, from the search for a resonance with mass $\MROne = \Mgkk = 2.9\TeV$  decaying to a second resonance with mass $\MRTwo = \Mphi = 0.4\TeV$ and a gluon. The figure shows the data (black points), the resulting background-only fit (solid line) and its uncertainty (barely visible gray hatched area), and the signal normalized to a cross section equal to the 95\% CL observed limit (dashed line). The data shown in each bin are the weighted sum of the number of events within each category, divided by the bin width, as a function of the dijet mass, with vertical bars representing the statistical uncertainty ($\sigma_{\text{stat}}$). The weight of each event is equal to the fraction of signal events in the category to which it is assigned, assuming a signal cross section equal to the 95\% \CL observed upper limit. The same quantities are also shown for the background-only fits in each category, and for the signal. The lower panel shows the difference between the data and the background prediction (points), and the background uncertainty (hatched gray area), divided by the statistical uncertainty.}\label{fig:fit_example}
\end{figure}

Figure~\ref{fig:fit_example} compares the dijet mass spectrum to the background-only fit, combined for the 22 categories of the signal hypothesis with $\Mgkk=2.9\TeV$ and $\Mphi=0.4\TeV$. In this combination, implemented for illustrative purposes, the events are weighted by the signal event fraction for each category, following the procedure of Ref.~\cite{Barlow:1987eaa}. These event fractions are calculated for \mjj values in a window of $\pm$20\% around the signal peak. We obtain the weights assuming a signal cross section equal to the observed 95\% \CL upper limit, and the resultant dijet mass distribution of this signal is also shown in Fig.~\ref{fig:fit_example}. The fit is displayed in the portion of the \mjj fit range common to all the categories. 

\begin{figure}[htb!]
	\centering                                             
    	\includegraphics[width=0.49\textwidth]{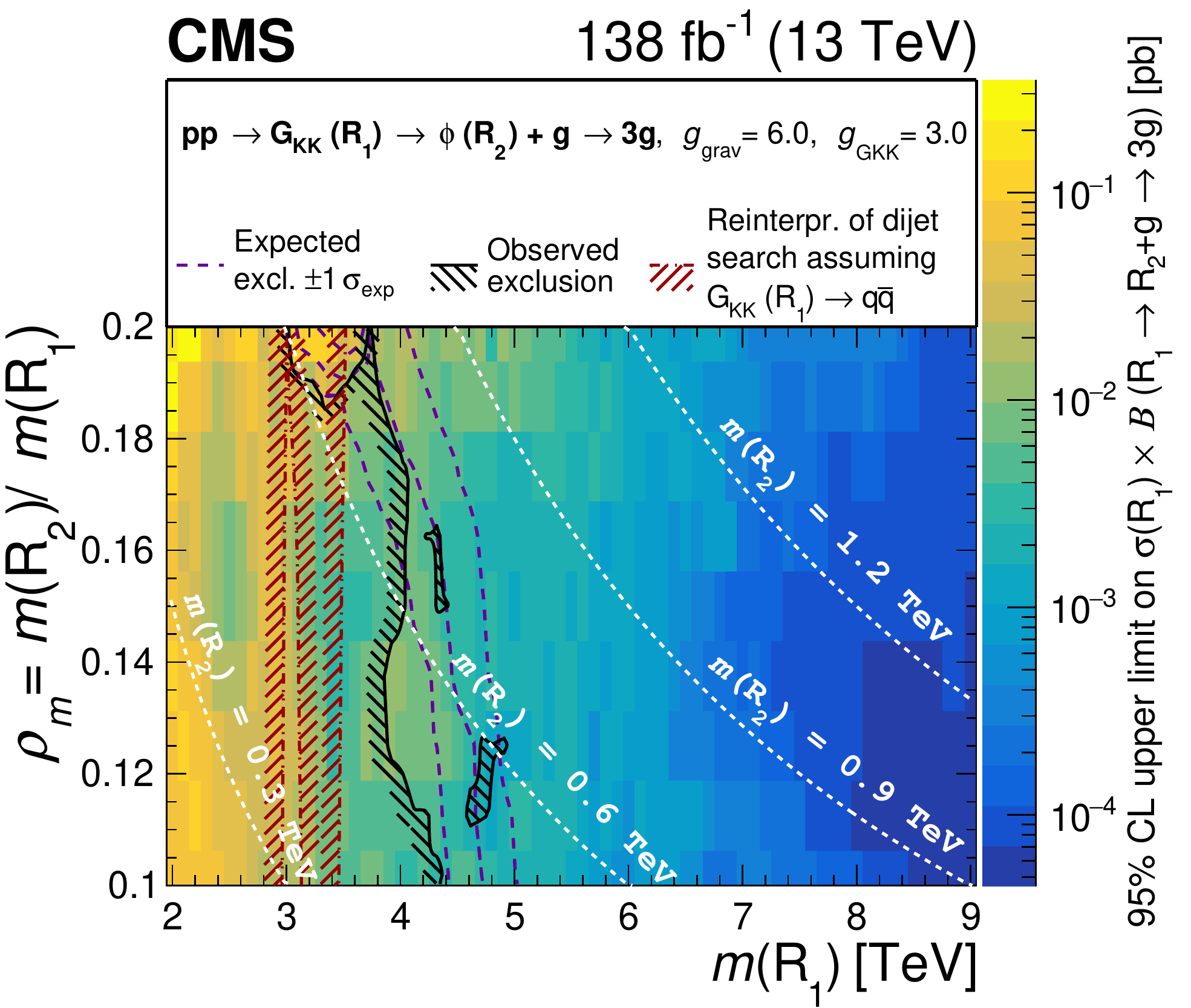} 
        	\caption{Observed upper limits on the product of signal cross section and branching fraction, as a function of \Ratio vs. \MROne, for a resonance model with three gluons in the final state. The excluded regions from this search (black hatched) are optimized for the $\gkk \to \phi + \Pg \to \Pg\Pg\Pg$ decay with $\gravCoupling = 6.0$ and $\gkkCoupling = 3.0$. These excluded regions are compared with those obtained from a reinterpretation of the inclusive CMS dijet resonance search (JHEP 05 (2020) 033, ~\cite{Sirunyan:2019vgj}), which is more sensitive to the decay channel $\gkk \to \PQq\PAQq$ (red hatched). The vertical band between the \MROne values of $\approx$3.0 and $\approx$3.1\TeV, for $\Ratio \lesssim 0.19$, is not excluded by the dijet search because of an upward statistical fluctuation in the observed limit. The white, dashed lines represent a sample of curves corresponding to fixed \MRTwo values.} \label{fig:limits}
\end{figure}

The modified frequentist \CLs criterion~\cite{Junk1999,Read_2002} is used to set upper limits on the signal cross section, following the prescription described in Ref.~\cite{CMS-NOTE-2011-005} using
the asymptotic approximation of the test statistic~\cite{Cowan:2010js}. Upper limits at 95\% \CL on the product of the production cross section of a \gkk and the $\mathcal{B}(\gkk \to \phi + \Pg \to \Pg\Pg\Pg)$ are derived for the different $\MROne=\Mgkk$ and $\Ratio=\Mphi/\Mgkk$ hypotheses. 
These limits, reported in Fig.~\ref{fig:limits}, are compared with the corresponding theoretical predictions for the cross section for a benchmark model with couplings $\gravCoupling=6$ and $\gkkCoupling=3$. For this choice of couplings the branching fraction of the decay $\gkk \to \phi + \Pg$ is between 50 and 60\% for all the signal hypotheses considered, while the rest of the decays are $\gkk \to \PQq\PAQq$. It can be seen that a wide range of resonance masses are excluded for the model. The dip in the expected and observed limit contours around $\MROne \approx 3.4\TeV$ and $\Ratio \approx 0.2$ is due to variations in the signal efficiency caused by the removal of categories in the fit, as described above.
The two isolated excluded regions, occurring in the \MROne interval between 4 and 5\TeV, are separated from the main excluded region at lower masses by a region where the observed limits  are higher than expected, consistent with an upward statistical fluctuation within this intermediate region.

The figure also shows the excluded region obtained from a reinterpretation of the inclusive CMS dijet resonance search~\cite{Sirunyan:2019vgj}, which is more sensitive to the decay channel $\gkk \to \PQq\PAQq$. For this reinterpretation we compared the theoretical cross section for $\PQq\PAQq \to \gkk \to \PQq\PAQq$ production, including all flavors of final state quarks except the top quark, with the observed upper limits from that search on the cross section for the production of a narrow resonance. Since the branching fraction of the \gkk to a quark-antiquark pair depends only weakly on the mass of the $\phi$ radion, the contours of the excluded area are approximately vertical in the figure. The two results shown in the figure represent the present CMS reach for this benchmark model of new physics in two independent decay channels. Constraints on the \gkk mass from searches in $\PQt\PAQt$ channels~\cite{CMS:2018rkg, ATLAS:2020lks} are comparable to those from the inclusive dijet analysis shown in Fig.~\ref{fig:limits}.

\section{Summary}\label{sec:summary}

A search for high-mass hadronic resonances that decay to a parton and a Lorentz-boosted resonance, which in turn decays into a pair of partons, has been presented. This is the first dedicated search for resonances decaying into three final state partons at the LHC in events with a boosted resonance. No statistically significant excess above the background predictions is observed. Results are interpreted in a model with a warped extra dimension where only the gluon field among the SM gauge fields is allowed to propagate in the entire bulk. The high-mass resonance is a Kaluza--Klein gluon, the boosted resonance is a radion, and the final-state partons are all gluons. Assuming this model, results from existing searches do not place constraints on the radion since it can only decay to a pair of gluons. By exploring a novel experimental signature, we significantly extend the excluded region in the parameter space of this benchmark model of new physics compared to previous inclusive searches for dijet resonances. In particular, the observed limits on the Kaluza–Klein gluon mass are extended by approximatively 1\TeV.

\begin{acknowledgments}
	We congratulate our colleagues in the CERN accelerator departments for the excellent performance of the LHC and thank the technical and administrative staffs at CERN and at other CMS institutes for their contributions to the success of the CMS effort. In addition, we gratefully acknowledge the computing centers and personnel of the Worldwide LHC Computing Grid and other centers for delivering so effectively the computing infrastructure essential to our analyses. Finally, we acknowledge the enduring support for the construction and operation of the LHC, the CMS detector, and the supporting computing infrastructure provided by the following funding agencies: BMBWF and FWF (Austria); FNRS and FWO (Belgium); CNPq, CAPES, FAPERJ, FAPERGS, and FAPESP (Brazil); MES and BNSF (Bulgaria); CERN; CAS, MoST, and NSFC (China); MINCIENCIAS (Colombia); MSES and CSF (Croatia); RIF (Cyprus); SENESCYT (Ecuador); MoER, ERC PUT and ERDF (Estonia); Academy of Finland, MEC, and HIP (Finland); CEA and CNRS/IN2P3 (France); BMBF, DFG, and HGF (Germany); GSRI (Greece); NKFIA (Hungary); DAE and DST (India); IPM (Iran); SFI (Ireland); INFN (Italy); MSIP and NRF (Republic of Korea); MES (Latvia); LAS (Lithuania); MOE and UM (Malaysia); BUAP, CINVESTAV, CONACYT, LNS, SEP, and UASLP-FAI (Mexico); MOS (Montenegro); MBIE (New Zealand); PAEC (Pakistan); MSHE and NSC (Poland); FCT (Portugal); JINR (Dubna); MON, RosAtom, RAS, RFBR, and NRC KI (Russia); MESTD (Serbia); MCIN/AEI and PCTI (Spain); MOSTR (Sri Lanka); Swiss Funding Agencies (Switzerland); MST (Taipei); ThEPCenter, IPST, STAR, and NSTDA (Thailand); TUBITAK and TAEK (Turkey); NASU (Ukraine); STFC (United Kingdom); DOE and NSF (USA).
		
	\hyphenation{Rachada-pisek} Individuals have received support from the Marie-Curie program and the European Research Council and Horizon 2020 Grant, contract Nos.\ 675440, 724704, 752730, 758316, 765710, 824093, 884104, and COST Action CA16108 (European Union); the Leventis Foundation; the Alfred P.\ Sloan Foundation; the Alexander von Humboldt Foundation; the Belgian Federal Science Policy Office; the Fonds pour la Formation \`a la Recherche dans l'Industrie et dans l'Agriculture (FRIA-Belgium); the Agentschap voor Innovatie door Wetenschap en Technologie (IWT-Belgium); the F.R.S.-FNRS and FWO (Belgium) under the ``Excellence of Science -- EOS" -- be.h project n.\ 30820817; the Beijing Municipal Science \& Technology Commission, No. Z191100007219010; the Ministry of Education, Youth and Sports (MEYS) of the Czech Republic; the Deutsche Forschungsgemeinschaft (DFG), under Germany's Excellence Strategy -- EXC 2121 ``Quantum Universe" -- 390833306, and under project number 400140256 - GRK2497; the Lend\"ulet (``Momentum") Program and the J\'anos Bolyai Research Scholarship of the Hungarian Academy of Sciences, the New National Excellence Program \'UNKP, the NKFIA research grants 123842, 123959, 124845, 124850, 125105, 128713, 128786, and 129058 (Hungary); the Council of Science and Industrial Research, India; the Latvian Council of Science; the Ministry of Science and Higher Education and the National Science Center, contracts Opus 2014/15/B/ST2/03998 and 2015/19/B/ST2/02861 (Poland); the Funda\c{c}\~ao para a Ci\^encia e a Tecnologia, grant CEECIND/01334/2018 (Portugal); the National Priorities Research Program by Qatar National Research Fund; the Ministry of Science and Higher Education, projects no. 14.W03.31.0026 and no. FSWW-2020-0008, and the Russian Foundation for Basic Research, project No.19-42-703014 (Russia); MCIN/AEI/10.13039/501100011033, ERDF ``a way of making Europe", and the Programa Estatal de Fomento de la Investigaci{\'o}n Cient{\'i}fica y T{\'e}cnica de Excelencia Mar\'{\i}a de Maeztu, grant MDM-2017-0765 and Programa Severo Ochoa del Principado de Asturias (Spain); the Stavros Niarchos Foundation (Greece); the Rachadapisek Sompot Fund for Postdoctoral Fellowship, Chulalongkorn University and the Chulalongkorn Academic into Its 2nd Century Project Advancement Project (Thailand); the Kavli Foundation; the Nvidia Corporation; the SuperMicro Corporation; the Welch Foundation, contract C-1845; and the Weston Havens Foundation (USA).
\end{acknowledgments}

\bibliography{auto_generated}
\cleardoublepage \appendix\section{The CMS Collaboration \label{app:collab}}\begin{sloppypar}\hyphenpenalty=5000\widowpenalty=500\clubpenalty=5000\cmsinstitute{Yerevan~Physics~Institute, Yerevan, Armenia}
A.~Tumasyan
\cmsinstitute{Institut~f\"{u}r~Hochenergiephysik, Vienna, Austria}
W.~Adam\cmsorcid{0000-0001-9099-4341}, J.W.~Andrejkovic, T.~Bergauer\cmsorcid{0000-0002-5786-0293}, S.~Chatterjee\cmsorcid{0000-0003-2660-0349}, M.~Dragicevic\cmsorcid{0000-0003-1967-6783}, A.~Escalante~Del~Valle\cmsorcid{0000-0002-9702-6359}, R.~Fr\"{u}hwirth\cmsAuthorMark{1}, M.~Jeitler\cmsAuthorMark{1}\cmsorcid{0000-0002-5141-9560}, N.~Krammer, L.~Lechner\cmsorcid{0000-0002-3065-1141}, D.~Liko, I.~Mikulec, P.~Paulitsch, F.M.~Pitters, J.~Schieck\cmsAuthorMark{1}\cmsorcid{0000-0002-1058-8093}, R.~Sch\"{o}fbeck\cmsorcid{0000-0002-2332-8784}, D.~Schwarz, S.~Templ\cmsorcid{0000-0003-3137-5692}, W.~Waltenberger\cmsorcid{0000-0002-6215-7228}, C.-E.~Wulz\cmsAuthorMark{1}\cmsorcid{0000-0001-9226-5812}
% ...
\vskip\cmsinstskip
% ...
1:~Also at TU Wien, Wien, Austria\\
% ...
\end{sloppypar}
\end{document}